\documentstyle[aps]{revtex}
\begin{document}
\draft
\title{Quantum Jumps on a Circle}
\author{K.\ Kowalski and  J.\ Rembieli\'nski}
\address{Department of Theoretical Physics, University
of \L\'od\'z, ul.\ Pomorska 149/153,\\ 90-236 \L\'od\'z,
Poland}
\maketitle
\begin{abstract}
It is demonstrated that in contrast to the well-known case with a
quantum particle moving freely in a real line, the wave packets
corresponding to the coherent states for a free quantum particle on
a circle do not spread but develop periodically in time.  The
discontinuous changes during the course of time in the phase
representing the position of a particle can be interpreted as the
quantum jumps on a circle.
\end{abstract}
\section*{}
In spite of the recent great progress in nanotechnology enabling for
example the construction of nanoscopic quantum rings \cite{1}, the
theory of quantization in the case when the configuration space
exhibits a nontrivial topology can hardly be called satisfactory.
For example, the coherent states for such a simple system as a
quantum particle on a circle have been introduced only recently
\cite{2,3}.  In this work we discuss the spreading of the wave packets
in the case of a free motion of a quantum particle in a circle.
Namely, we show that in contrast to the popular example of a free
evolution on a real line \cite{4} the wave packets referring to the
coherent states for a quantum particle on a circle do not spread
over the configuration space but execute regular oscillations.  We
also demonstrate that such form of the evolution of the wave packets
leads to the discontinuous changes in the expectation values of the
phase operator i.e.\ the position observable for a quantum particle
on a circle.  From the classical point of view such discontinuities
can be viewed as the quantum jumps.

We first collect the basic facts about the quantum mechanics on a
circle $S^1$.  Consider a free quantum particle on $S^1$.  We assume for
simplicity that the particle has unit mass and it moves in a unit
circle.  The classical Hamiltonian is of the form
\begin{equation}
H = \hbox{$\frac{1}{2}$}J^2,
\end{equation}
where $J$ is the angular momentum canonically conjugated to the
angle $\varphi$ i.e.\ we have the Poissson bracket such that
\begin{equation}
\{\varphi,J\} = 1,
\end{equation}
leading, accordingly to the rules of the canonical quantization to the
commutator
\begin{equation}
[\hat\varphi,\hat J] = i,
\end{equation}
where we set $\hbar=1$.  Motivated by numerous misinterpretations
and misunderstandings which can be met in the literature related
with the commutator (3) we now study (3) in a more detail.  Consider
the Hilbert space ${\cal H}$ of square integrable functions on $S^1$ 
satisfying the generalized periodicity condition $f(\varphi+2\pi)~=~
e^{2\pi{\rm i}\lambda}f(\varphi)$, where $\lambda$ is fixed and 
$0\le\lambda<1$, specified by the scalar product
\begin{equation}
\langle f,g\rangle =\frac{1}{2\pi}\int_{0}^{2\pi}d\varphi
f^*(\varphi)g(\varphi).
\end{equation}
In the following we shall call the elements of ${\cal H}$ the
$\lambda$-phase periodic functions.  The integral (4) is understood as 
the Lebesgue integral.  It is also important to note that we identify 
the vectors in ${\cal H}$ equal to one another almost everywhere.  More 
precisely, denoting the set of $\lambda$-phase periodic functions with 
the domain having Lebesgue measure zero by $N$, the functions $f$ and 
$f+N$ represent the same vector (equivalence class) in ${\cal H}$, i.e.\ we 
actually have ${\cal H}=L^2(S^1)/N$.  In the following we will represent 
vectors in ${\cal H}$ by their representatives.  Now, the hermitian angular 
momentum operator $\hat J$ acts on the differentiable functions (representatives 
of vectors belonging to the domain of $\hat J$) in the following way:
\begin{equation}
\hat J f(\varphi) = -{\rm i}\frac{d}{d\varphi}f(\varphi).
\end{equation}
We remark that there exists the selfadjoint closure of $\hat J$
defined on the set of $\lambda$-phase periodic absolutely continuous functions.
Of course, the operator of the multiplication by $\varphi$, suggested by  
the naive utilization of (3) and (5), maps a $\lambda$-phase periodic function 
into non-$\lambda$-phase periodic function.  The only exception is the zero 
function $f(\varphi)\equiv0$.  An adequate selfadjoint bounded angle operator 
preserving the $\lambda$-phase periodicity of a function is defined by
\begin{equation}
\hat \varphi f(\varphi)=\left(\varphi
-2\pi\left[\frac{\varphi}{2\pi}\right]\right)f(\varphi),
\end{equation}
where $[x]$ is the biggest integer in $x$.  We remark that an
analogous form of the angle operator was indicated in \cite{5}.  Now,
it is evident that the commutator of $\hat J$ and $\hat\varphi$ is
trivially defined in ${\cal H}$.  Indeed, for arbitrary nontrivial
$\lambda$-phase periodic function $f\in {\cal H}$, $\hat\varphi f$ is an
element of ${\cal H}$ but due to the discontinuity of $\hat\varphi f$
resulting from (6) it is not an element of the domain of $\hat J$.
We conclude that the canonical commutation relation (3) is defined only 
on the zero vector.  It seems that the most natural solution of this 
problem is to introduce the unitary operator
\begin{equation}
U = e^{i\hat\varphi}
\end{equation}
satisfying
\begin{equation}
Uf(\varphi)=e^{{\rm i}\hat\varphi}f(\varphi)=e^{{\rm i}\left(\varphi
-2\pi\left[\frac{\varphi}{2\pi}\right]\right)}f(\varphi)=e^{{\rm
i}\varphi}f(\varphi),
\end{equation}
where $\hat\varphi$ is given by (6).  Now, an immediate consequence of
(5) and (8) is the following commutation relation:
\begin{equation}
[\hat J,U] = U.
\end{equation}
Since $U$ is defined on an arbitrary element of ${\cal H}$ and it
leaves invariant the domain of $\hat J$, therefore, in opposition to (5), 
the commutation relation (9) is defined on the dense set in ${\cal H}$.  
This observation as well as behavior of the expectation values of 
$U$ in the coherent state (see figures 1 and 3) suggest that $U$ 
is a better candidate to represent the position of a quantum particle 
on a circle than $\hat\varphi$.  The unitary operator $U$ satisfying
(9) was discussed for the first time by Luisell \cite{6}.  We stress 
that the problem of the description of the angular position should not 
be mixed up with the problem of the proper definition of the phase operator
related to the polar decomposition of the Bose annihilation operator
for harmonic oscillator \cite{7}.

Now, let $e_j(\varphi)=e^{{\rm i}j\varphi}$, where $j=k+\lambda$, and
$k$ is integer, represent the eigenvectors of the angle momentum operator 
$\hat J$
\begin{equation}
\hat J e_j(\varphi) = j e_j(\varphi)
\end{equation}
corresponding to the eigenvalue $j$.  The operators $U$ and
$U^\dagger$ act on the vectors $e_j(\varphi)$ as the ladder
operators.  Namely
\begin{mathletters}
\begin{eqnarray}
Ue_j(\varphi) &=& e_{j+1}(\varphi),\\
U^\dagger e_j(\varphi) &=& e_{j-1}(\varphi).
\end{eqnarray}
\end{mathletters}
Demanding the time-reversal invariance of the algebra (9) we find that
the only possibility left is $\lambda=0$ or $\lambda=1/2$ \cite{2}.  Therefore, $j$
can be only integer or half-integer.  We refer to the case with
integer (half-integer) $j$ as to the boson (fermion) case.  Clearly, the
vectors $e_j(\varphi)$ span the Hilbert space of states ${\cal H}$.
We finally point out that the case with $\lambda=0$ ($\lambda=1/2$) refers to
periodic (anti-periodic) functions.

Consider now the coherent states for a particle on a circle
\cite{2}.  These states can be defined as a solution of the
eigenvalue equation
\begin{equation}
Zf_\xi (\varphi) = \xi f_\xi (\varphi),
\end{equation}
where $Z=e^{-\hat J+\frac{1}{2}}U$, and the complex number
$\xi=e^{-l +{\rm i}\alpha}$ parametrizes the cylinder
which is the classical phase space for the particle moving in a circle.
The coherent states are given by
\begin{mathletters}
\begin{eqnarray}
f_\xi(\varphi)  &=&
\theta_3(\hbox{$\frac{1}{2\pi}$}(\varphi-\alpha-{\rm i}l)
|\hbox{$\frac{{\rm i}}{2\pi}$}),\qquad \hbox{(boson case)}\\
f_\xi(\varphi) &=&
\theta_2(\hbox{$\frac{1}{2\pi}$}(\varphi-\alpha-{\rm i}l)
|\hbox{$\frac{{\rm i}}{2\pi}$}),\qquad \hbox{(fermion case)}
\end{eqnarray}
\end{mathletters}
where $\theta_3$ and $\theta_2$ are the Jacobi theta-functions defined by
\begin{mathletters}
\begin{eqnarray}
\theta_3(v|\tau) &=&
\sum_{n=-\infty}^{\infty}q^{n^2}(e^{i\pi v})^{2n},\\
\theta_2(v|\tau) &=&
\sum_{n=-\infty}^{\infty}q^{(n-\frac{1}{2})^2}(e^{i\pi
v})^{2n-1},
\end{eqnarray}
\end{mathletters}
where $q=e^{i\pi\tau}$ and $\hbox{Im}\,\tau>0$.  It is interesting
that the above coherent states have the Bargmann representation
which is closely related to the general construction introduced by 
Stenzel \cite{8}.
The overlap integral is
\begin{mathletters}
\begin{eqnarray}
\langle f_\xi, f_{\xi'}\rangle  &=&
\theta_3(\hbox{$\frac{1}{2\pi}$}(\alpha-\alpha')-\hbox{$\frac{
l+l'}{2}$}\hbox{$\frac{{\rm i}}{\pi}$}
|\hbox{$\frac{{\rm i}}{\pi}$}),\qquad \hbox{(boson case)}\\
\langle f_\xi, f_{\xi'}\rangle  &=&
\theta_2(\hbox{$\frac{1}{2\pi}$}(\alpha-\alpha')-\hbox{$\frac{
l+l'}{2}$}\hbox{$\frac{{\rm i}}{\pi}$}
|\hbox{$\frac{{\rm i}}{\pi}$}).\qquad \hbox{(fermion case)}
\end{eqnarray}
\end{mathletters}
The expectation value $\langle\hat J\rangle_{f_\xi }$ of $\hat J$ in
the normalized coherent state $f_\xi/\|f_\xi\|$ is
\begin{equation}
\langle\hat J\rangle_{f_\xi } \approx l,
\end{equation}
where we have the exact equality in the case with $l$ integer or 
half-integer and the maximal error arising in the case $l\to0$ is of order
$0.1$\%.  Therefore, the parameter $l$ in $\xi$ can be really
identified with the classical angular momentum.  Further, we have
the following formula on the relative expectation value $\langle
U\rangle_{f_\xi}/\langle U\rangle_{f_1}$, which is the most natural
candidate to describe the average position of a particle on a circle:
\begin{equation}
\frac{\langle U\rangle_{f_\xi}}{\langle U\rangle_{f_1}}\approx
e^{{\rm i}\alpha},
\end{equation}
where the approximation is very good.  We conclude that
the parameter $\alpha$ can be interpreted as a classical
angle.

We now specialize to free motion in a circle.  Clearly, the quantum
Hamiltonian is
\begin{equation}
\hat H = \hbox{$\frac{1}{2}$}\hat J^2.
\end{equation}
It turns out that as with the standard coherent states the discussed
coherent states for a particle on a circle are not stable with
respect to the free evolution.  Namely, we have
\begin{equation}
Z(t)f_\xi(\varphi) = e^{\frac{-{\rm i}t}{2}}\xi f_\xi (\varphi +t),
\end{equation}
where $Z(t)=e^{{\rm i}t\frac{{\hat J}^2}{2}}Ze^{-{\rm i}t\frac{{\hat J}^2}
{2}}=e^{{\rm i}t(\hat J-\frac{1}{2})}Z$.  Therefore, the problem
naturally arises as to whether the wave packet $f_\xi(\varphi)$ spreads
in the circle as in the standard case of the free motion in the real
line.  Consider the probability distribution in the coordinate space.
As time develops the wave packet for a free particle on a circle is
given by
\begin{mathletters}
\begin{eqnarray}
f_\xi(\varphi,t)  &=& e^{-{\rm i}t\frac{{\hat J}^2}{2}}f_\xi(\varphi) =
\theta_3(\hbox{$\frac{1}{2\pi}$}(\varphi-\alpha-{\rm i}l)
|\hbox{$\frac{1}{2\pi}({\rm i}-t)$}),\qquad \hbox{(boson case)}\\
f_\xi(\varphi,t) &=& e^{-{\rm i}t\frac{{\hat J}^2}{2}}f_\xi(\varphi) =
\theta_2(\hbox{$\frac{1}{2\pi}$}(\varphi-\alpha-{\rm i}l)
|\hbox{$\frac{1}{2\pi}({\rm i}-t)$}).\qquad \hbox{(fermion case)}
\end{eqnarray}
\end{mathletters}
It can be easily checked that both the functions (20) are $4\pi$-
periodic.  Using (20) and (15) we find that the probability density for the
coordinates at time $t$ is
\begin{mathletters}
\begin{eqnarray}
p_{l,\alpha}(\varphi,t)  &=&
\frac{|f_\xi(\varphi,t)|^2}{\|f_\xi\|^2} =
\frac{|\theta_3(\hbox{$\frac{1}{2\pi}$}(\varphi-\alpha-{\rm i}l)
|\hbox{$\frac{1}{2\pi}({\rm i}-t)$})|^2}{\theta_3(\hbox{$\frac{{\rm
i}l}{\pi}$}|\hbox{$\frac{{\rm i}}{\pi}$})},\qquad \hbox{(boson case)}\\
p_{l,\alpha}(\varphi,t) &=&
\frac{|f_\xi(\varphi,t)|^2}{\|f_\xi\|^2} =
\frac{|\theta_2(\hbox{$\frac{1}{2\pi}$}(\varphi-\alpha-{\rm i}l)
|\hbox{$\frac{1}{2\pi}({\rm i}-t)$})|^2}{\theta_2(\hbox{$\frac{{\rm
i}l}{\pi}$}|\hbox{$\frac{{\rm i}}{\pi}$})}.\qquad \hbox{(fermion case)}
\end{eqnarray}
\end{mathletters}
The above probability densities are periodic functions of time
with the period $4\pi$ (boson case) and $2\pi$ (fermion case),
respectively.  It thus appears that in opposition to the standard
case of a free particle on the real line the wave packets do not
spread but behave rather like (linear) solitons.  As a consequence
of the oscillations of the probability density in the coordinate
space an interesting phenomenon occurs which can be interpreted as
quantum jumps on the circle.  Namely, it turns out that the
probability density has at $t=t_*=(2k+1)\pi$, where $k$ is
integer, two identical maxima (see Fig.\ 1).  Thus at $t=t_*$ the particle 
can be detected with equal (maximal) probabilities at two different points
on a circle.  What about the position on a circle at $t=t_*$ ?  In
order to answer this question we should first identify a classical
counterpart of the angle coordinate.  As one would expect having in
mind the experience with the commutator (3) the expectation value of
the angle operator in the Heisenberg picture $\hat \varphi(t)$ in the 
normalized coherent state can hardly be interpreted as such a counterpart 
(see Fig.\ 2).  Instead, it appears that an appropriate candidate
is the argument of the expectation value of the operator $U(t)$ (see
(7)) in the normalized coherent state $f_\xi/\|f_\xi \|$, so
\begin{equation}
\varphi_{{\rm class}}(t) = {\rm Arg}\langle U(t)\rangle_{f_\xi}\quad\hbox{{\rm mod} $2\pi$}.
\end{equation}
Using (22) and the formula
\begin{mathletters}
\begin{eqnarray}
\langle U(t)\rangle_{f_\xi}  &=& e^{-\frac{1}{4}}e^{{\rm i}\alpha }
\frac{\theta_2(\hbox{$\frac{t}{2\pi}-\frac{{\rm i}l}{\pi}$}
|\hbox{$\frac{{\rm i}}{\pi}$})}{\theta_3(\hbox{$\frac{{\rm
i}l}{\pi}$}|\hbox{$\frac{{\rm i}}{\pi}$})},\qquad \hbox{(boson case)}\\
\langle U(t)\rangle_{f_\xi}  &=& e^{-\frac{1}{4}}e^{{\rm i}\alpha }
\frac{\theta_3(\hbox{$\frac{t}{2\pi}-\frac{{\rm i}l}{\pi}$}
|\hbox{$\frac{{\rm i}}{\pi}$})}{\theta_2(\hbox{$\frac{{\rm
i}l}{\pi}$}|\hbox{$\frac{{\rm i}}{\pi}$})},\qquad \hbox{(fermion case)}
\end{eqnarray}
\end{mathletters}
where $U(t)=e^{{\rm i}t\frac{{\hat J}^2}{2}}Ue^{-{\rm i}t\frac{{\hat J}^2}
{2}}$, we find that there is a discontinuity in the angle at $t=t_*$
(see Fig.\ 3).  The limit approached from the left and the limit
approached from the right coincide with the abscissae of the two identical 
maxima of the probability density.  It seems plausible to interpret such
discontinuities as a manifestation of the quantum jumps on a circle.
We point out that the quantum jumps take place in the boson case
only for $l$ integer and in the fermion case only for $l$
half-integer.

Concluding, the wave packets for a particle moving
freely in a circle are periodic functions of time and do not spread.
As a consequence the discontinuity appears in the phase representing
the position on a circle which can be regarded as quantum jumps.  To
our knowledge this observation provides the first example of exotic
constrained dynamics of a quantum particle.

\newpage
\noindent{\bf Figure captions}
\begin{figure}
\caption{The probability density $p_{l,\alpha}$
given by (21a) (boson case), where $\alpha=0.75\pi,\,l=1$,
and $t=\pi$ (solid line).  The two maxima of the probability
density with the abscissa $\varphi=2.356$ and $\varphi=5.498$, respectively,
are identical and equal to 1.773.  For the better illustration of the
evolution of the probability density the curves are placed in 
the figure referring to $t=0.9\pi$ (dotted line) and $t=1.1\pi$ 
(dash line).}
\label{fig1}
\caption{The evolution of the expectation value of the angle
operator $\hat\varphi(t)$ given by $\langle\hat\varphi(t)\rangle$=
$\frac{1}{2\pi}\int_{0}^{2\pi}\varphi p_{l,\alpha}
(\varphi,t)d\varphi$, where the probability density $p_{l,\alpha}$ 
is given by (21a) (boson case); $\alpha$ and 
$l$ are the same as in Fig.\ 1.  It should be noted that $\langle
\hat \varphi(t)\rangle$ takes the values only from the subset of the 
circle $[0,2\pi)$ given approximately by the interval (1.8,4.8).
Further, the plot is not piecewise linear as one would expect for the
case of the free evolution on a circle.  Such behavior of $\langle\hat
\varphi(t)\rangle$ indicates that it is rather poor candidate to represent 
the classical angle.}
\label{fig2}
\caption{The time dependence of the counterpart of the classical
angle specified by (22) and (23a) (boson case).  The values ${\rm Arg}
\langle U(t)\rangle_{f_\xi}=2.356$ and ${\rm Arg}\langle
U(t)\rangle_{f_\xi}=5.498$, where the discontinuities appear, coincide with
the abscissa $\varphi=2.356$ and $\varphi=5.498$, respectively, of the two 
identical maxima of the probability density shown in Fig.\ 1.}
\end{figure}
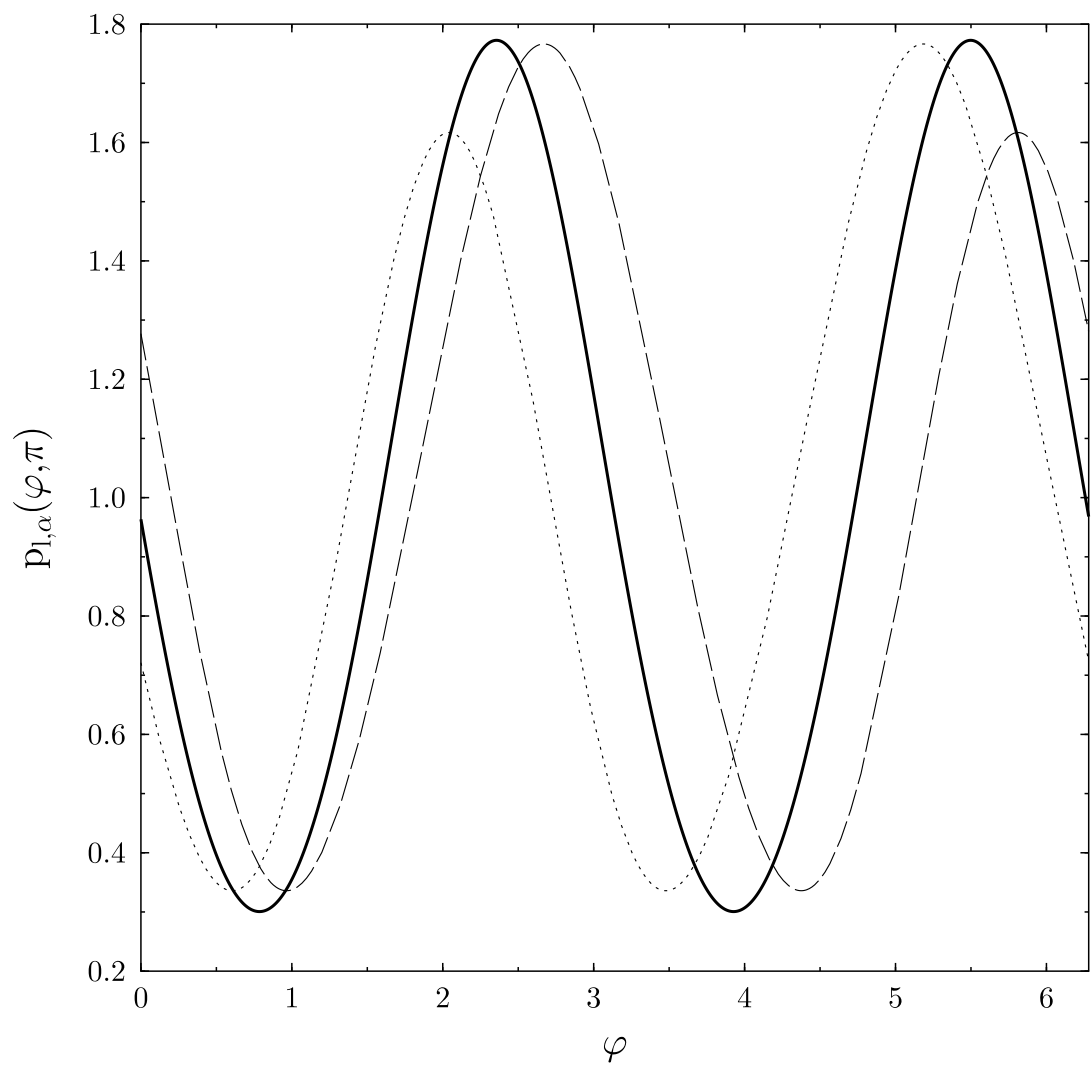
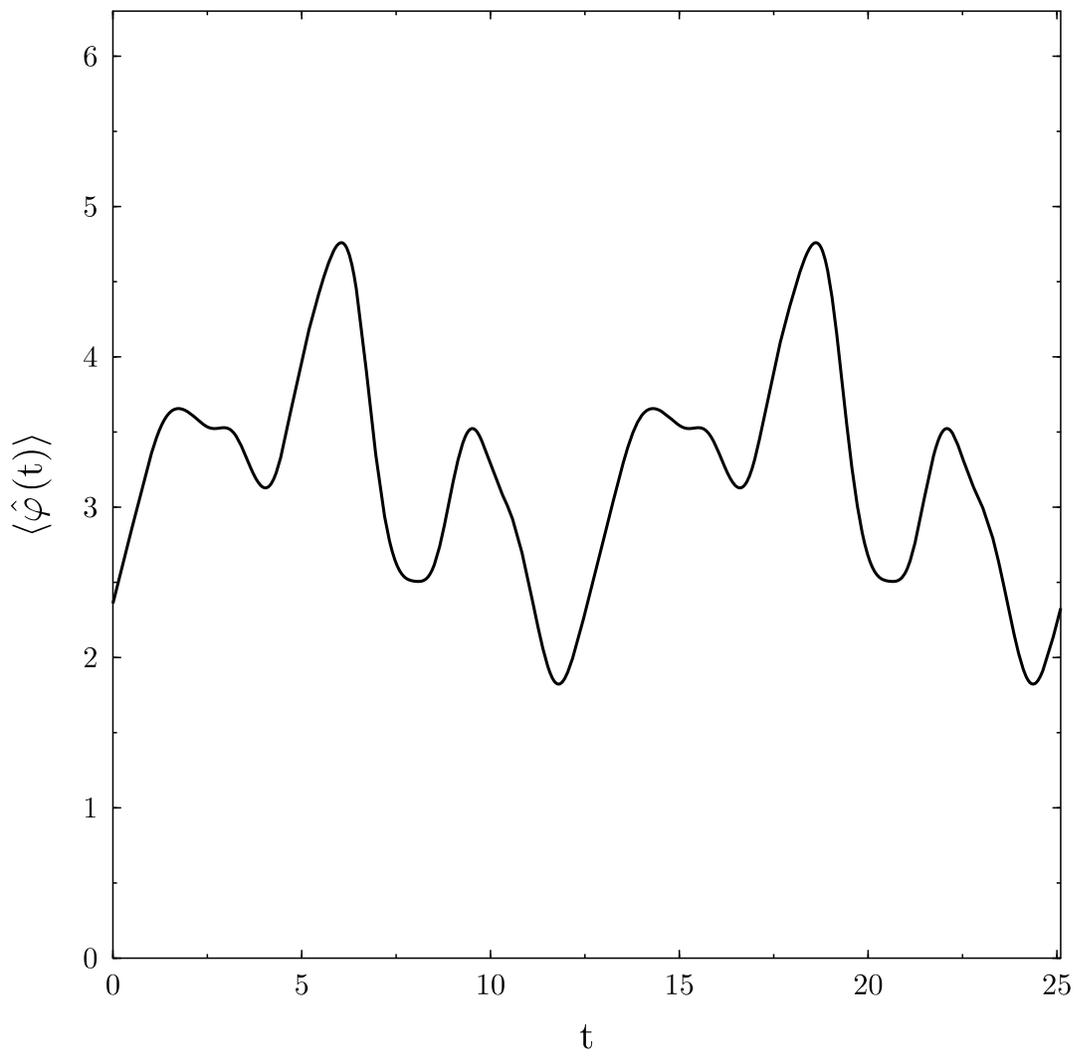
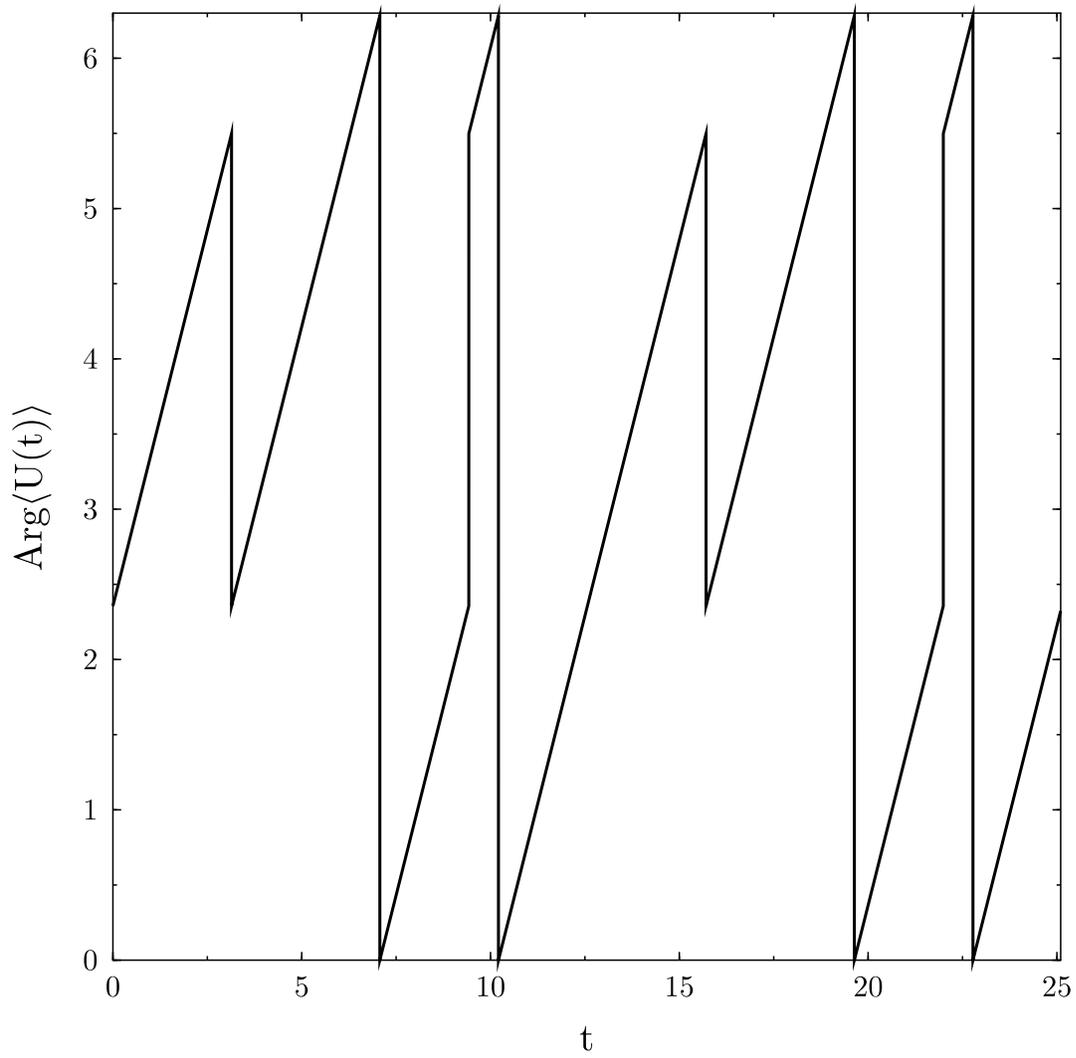
\end{document}